\documentclass[noshowpacs,amsmath,
twocolumn,
superscriptaddress,
8pt
]{revtex4-1}
\usepackage[utf8]{inputenc}
\usepackage{setspace}
\usepackage{amsmath}
\usepackage{float}
\usepackage{bm}
\usepackage{ulem}
\usepackage{graphicx}
\usepackage[nearskip,margin = 0pt]{subfig}

\usepackage{verbatim}
\usepackage{amsfonts}
\usepackage{braket}
\usepackage{amssymb}
\usepackage{upgreek}
\usepackage[colorlinks,linkcolor=blue,anchorcolor=blue,citecolor=blue,urlcolor=black]{hyperref}
\usepackage{epstopdf}
\usepackage{xcolor}
\usepackage{booktabs}
\usepackage{tabularx}
\usepackage{xtab}
\usepackage{changepage}
\usepackage{ragged2e}

\DeclareGraphicsExtensions{.pdf,.eps,.png,.jpg,.mps} 

\begin{document}

\title{Polarized electroluminescence with magnetic spectral tuning in van der Waals magnet CrSBr}
\author{Yilei Wang$^{1,2,\dagger}$, Shiqi Yang$^{1,\dagger,\star}$, Leyan Huang$^{1}$, Yuqia Ran$^{1}$, Pingfan Gu$^{3}$, Xinyue Huang$^{1,2}$, Kenji Watanabe$^{4}$, Takashi Taniguchi$^{5}$, Zuxin Chen$^{6,\star}$ and Yu Ye$^{1,7,8,\star}$\\
\vspace{6pt}
$^1$State Key Laboratory for Mesoscopic Physics and Frontiers Science Center for Nano-optoelectronics, School of Physics, Peking University, Beijing 100871, China\\
$^2$Academy for Advanced Interdisciplinary Studies, Peking University, Beijing 100871, China\\
$^3$MIIT Key Laboratory of Semiconductor Microstructure and Quantum Sensing Department of Applied Physics, Nanjing University of Science and Technology, Nanjing 210094, China.\\
$^4$Research Center for Electronic and Optical Materials, National Institute for Materials Science, 1-1 Namiki, Tsukuba 305-0044, Japan\\
$^5$Research Center for Materials Nanoarchitectonics, National Institute for Materials Science, 1-1 Namiki, Tsukuba 305-0044, Japan\\
$^6$School of Semiconductor Science and Technology, South China Normal University, Foshan 528225, China\\
$^7$Yangtze Delta Institute of Optoelectronics, Peking University, Nantong 226010 Jiangsu, China\\
$^8$Liaoning Academy of Materials, Shenyang 110167, China\\
\vspace{3pt}
$^{\dagger}$These authors contributed equally to this work.\\
$^{\star}$Corresponding to: yang\_shiqi@pku.edu.cn; chenzuxin@m.scnu.edu.cn; ye\_yu@pku.edu.cn}

\begin{abstract}
\begin{adjustwidth}{-2cm}{0cm}
\textbf{ABSTRACT:}
Polarized wavelength-tunable electroluminescence (EL) represents a critical on-demand functionality for next-generation optoelectronics. While conventional van der Waals (vdW) EL devices offer discrete wavelength switching constrained by fixed emission states, we report a novel platform enabling continuous spectral tuning combined with intrinsically polarized emission. By leveraging exciton-assisted inelastic tunneling in the anisotropic magnet CrSBr, our devices achieve uniform EL with a near unity degree of linear polarization ($\approx$ 94.3$\%$). The strong magneto-electronic coupling in CrSBr facilitates continuous magnetic-field-controlled spectral tuning through spin canting-induced band renormalization. This work establishes vdW magnets as a versatile platform for developing reconfigurable polarized light sources with simultaneous spectral and polarization control.

\end{adjustwidth}
\end{abstract}
\date{\today}
\maketitle

\noindent

Near-infrared (NIR) electroluminescence (EL) has attracted significant interest for applications spanning telecommunications, biological imaging, and quantum technologies\cite{2010NIR-telecom,2017NIR,2023NIRnature,2024NIRNN}. Particularly desirable is wavelength-tunable and polarized NIR light emission—a functionality demanding synchronous control over spectral position and polarization states. Two-dimensional (2D) materials have attracted widespread attention, as their electronic properties can be easily manipulated by composition\cite{pu2022continuous-alloying,duan2016synthesis-alloying}, strain\cite{2020strain,2022strain-tunable,2023strain}, and electric fields\cite{henck2022light,joe2021electrically,ross2014electrically,shin2024electrically,wang2019polarity,z2021AC-driven}. Numerous studies have concentrated on the monolayer transition metal dichalcogenides (TMDCs) due to the rich and complex excitonic scenario in the NIR region. In $\mathrm{WSe_2}$, for instance, tunable light emission switching from multi-particle exciton complexes has been achieved by electrostatic doping\cite{2019NM-WSe2}. However, most semiconductor EL devices demonstrated so far can switch only within a few fixed emission wavelengths determined by the energy of the exciton complexes states without polarization control. Anisotropic 2D materials like ReS$_2$ offer polarization control\cite{2020Res2} but lack continuous spectral tunability.

Chromium Sulfur Bromide (CrSBr), an air-stable vdW magnet\cite{wang2022air-stable}, emerges as an ideal candidate to bridge this gap. Its strong spin-charge coupling enables magnetic-field-driven band engineering\cite{2020Magnetic,2025Magnetic}, while extreme in-plane electronic anisotropy (quasi-1D character) intrinsically generates linealy polarized excitons\cite{2022anisotropy,qian2023anisotropic,2025Raman}. Recent photoluminescence (PL) studies confirm robust magneto-excitonic coupling—where spin canting under magnetic fields continuously modulates exciton energies—and near-perfect optical polarization along the crystallographic $b$-axis\cite{wu2024magneto,wilson2021interlayer,2025Mag}. Despite these advantages, CrSBr's potential for electrically driven polarized emission with magnetic tunability remains unexplored.

In this study, we develop an EL platform integrating three breakthroughs: (1) Exciton-assisted tunneling generates polarized EL with a degree of linear polarization (DOLP) $\approx$94.3$\%$ without external polarizers, (2) Magnetic fields continuously tune emission wavelengths over 8 meV via spin canting, and (3) Ultralow operating currents enabled by efficient energy transfer. This synergy between intrinsic polarization and magnetic reconfigurability opens pathways toward multifunctional on-chip light sources.

\begin{figure*}[tbh]    
\includegraphics[width=2\columnwidth]{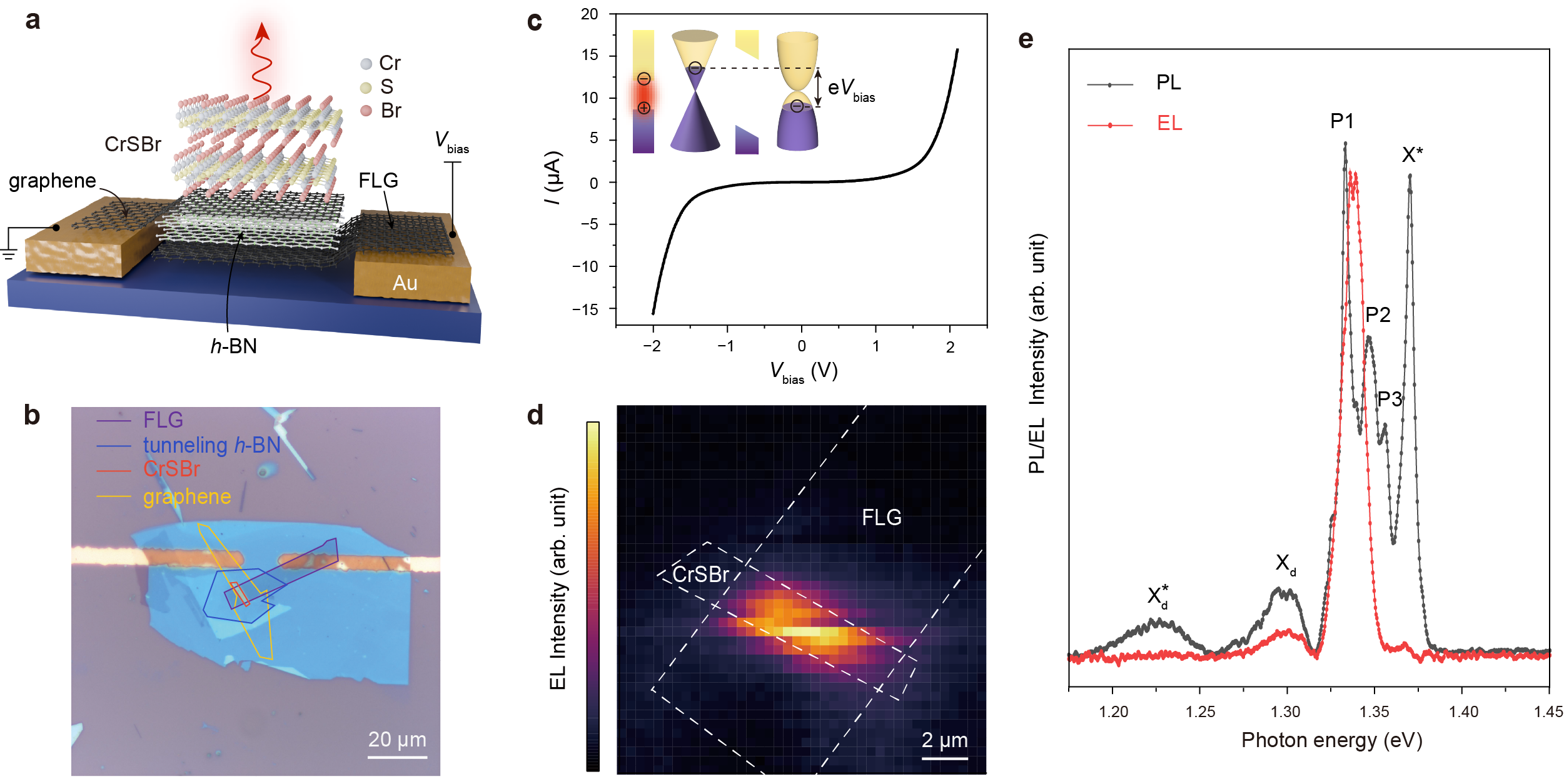}
\captionsetup{singlelinecheck=off, justification = RaggedRight}
\caption{\label{F1}\textbf{Electroluminescent device based on exciton-assisted tunneling vdW heterostructure.}
     \textbf{(a)} Schematic illustration of CrSBr electroluminescent device.
     \textbf{(b)} Optical microscopy image of Device 1. The yellow, purple, orange, and blue solid lines outline the graphene, FLG, few-layer CrSBr, and tunneling \textit{h}-BN layer, respectively.
     \textbf{(c)} Current-voltage characteristic of the device at 2 K. Inset: Band diagram under a positive bias $V_{\rm{bias}}$, illustrating the energy levels of CrSBr, graphene, \textit{h}-BN, and FLG (from left to right). Electrons tunnel from graphene to FLG inelastically, assisting the generation of excitons (encircled in CrSBr layers).
     \textbf{(d)} Spacial mapping of EL intensity from Device 1 at an applied bias of 2.1 V. The electroluminescence is confined to the region where the CrSBr flake overlaps with the tunneling junction.
     \textbf{(e)} Comparison of PL (black curve) and EL (red curve) spectra at 2 K. The EL spectrum was acquired at 2.1 V. The PL spectrum shows multiple peaks corresponding to surface (P) and bulk ($\mathrm{X}^*$) excitons, while the EL spectrum is dominated by the surface exciton peak (P).
}
\end{figure*}

Our electroluminescence devices are based on an exciton-assisted tunneling mechanism\cite{2023exciton-assited}, with a heterostructure comprising a graphene electrode, hexagonal boron nitride (\textit{h}-BN) tunnel barrier layer (2.5 nm in thickness), and few-layer graphene (FLG). A thin flake of CrSBr (15 nm, Device 1) was positioned adjacent to the tunneling pathway, as illustrated in Figure \ref{F1}a. The vertical heterostructure was fabricated using mechanical exfoliation followed by a dry-transfer technique, with an optical image of the completed device shown in Figure \ref{F1}b. 

During operation, the graphene electrode was grounded while a bias voltage ($V_{\text{bias}}$) was applied to the FLG electrode. The device exhibited characteristic tunneling behavior, as shown in the current-voltage ($I$–$V_{\text{bias}}$) curve (Figure \ref{F1}c), which adheres to the Fowler–Nordheim tunneling model\cite{1969Fowler,2018F-N} (Figure S1). When the applied positive voltage exceeds the threshold, inelastic electron tunneling from graphene to FLG facilitates energy transfer to the adjacent CrSBr layer, generating excitons that undergo radiative recombination and produce EL (inset of Figure \ref{F1}c).

Theoretically, light emission should occur once the energy transfer condition is satisfied (i.e., e$V_{\text{bias}}$\textgreater1.34 eV). However, detectable EL signals from Device 1 were only observed at bias voltages above 1.5 V (corresponding to a current of 2.24 $\upmu$A; Figure S2). This deviation likely arises from the low tunneling probability at lower bias voltages, leading to EL intensity below the spectrometer's detection limit. Furthermore, interfacial effects may reduce energy transfer efficiency, necessitating higher threshold voltages for observable emission. Spatial EL mapping at 2.1 V (15.9 $\upmu$A) confirms uniform emission localized within the heterostructure's overlap region (Figure \ref{F1}d).

\begin{figure*}[tbh]    
\includegraphics[width=1.7\columnwidth]{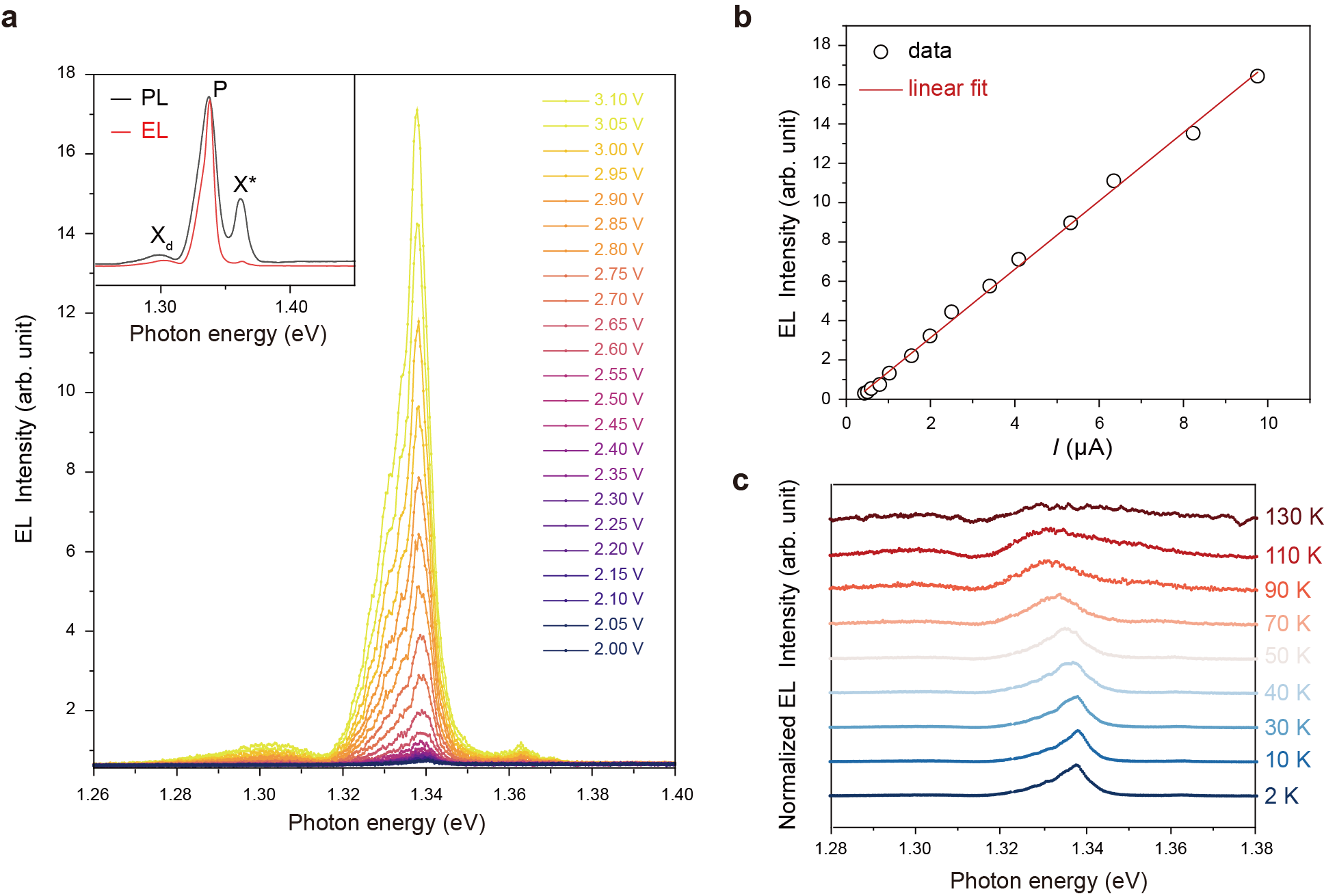}
\captionsetup{singlelinecheck=off, justification = RaggedRight}
\caption{\label{F2}\textbf{EL characteristics of Device 2.}
     \textbf{(a)} EL spectra at different bias voltages measured at 2 K. Inset: EL spectrum (red) at 3.1 V compared to the PL spectrum (black).
     \textbf{(b)} EL intensity of the P peak as a function of tunneling current at 2 K. The red line indicates a linear fit.
     \textbf{(c)} Temperature evolution of normalized EL spectra from 2 K to 130 K at a fixed bias of 3.1 V. The P peak redshifts with increasing temperature up to 90 K and broadens significantly; the emission quenches above 110 K.
}
\end{figure*}

Few-layer CrSBr exhibits two distinct exciton components: surface and bulk excitons\cite{2025Mag}. The PL spectrum of Device 1 (Figure \ref{F1}e) shows multiple peaks, consistent with prior studies on 15 nm CrSBr flakes\cite{2024band-split,2023ACSinterplay}. In thicker CrSBr flakes, the surface exciton peak (P) broadens and splits into three resolved components (P1-P3)\cite{2023ACSinterplay}, while the $\mathrm{X}^*$ peak represents the bulk exciton. The low-energy peaks ($\mathrm{X}_d$ and $\mathrm{X}_d^*$) arise from defect-related states\cite{2022defect}. For the 4 nm CrSBr in Device 2, PL reveals only $\mathrm{X}_d$, $\mathrm{X}^*$, and a single P peak. Interestingly, the EL from both devices shows fewer peaks, with just a dominant P exciton and weak $\mathrm{X}_d$ feature (Figure \ref{F1}e and Figure \ref{F2}a inset). Accordingly, Device 2's EL P peak displays narrower linewidth than its PL counterpart, likely because PL probes surface excitons in asymmetric dielectric environments (top \textit{h}-BN vs. bottom graphene), while EL selectively excites only the graphene-proximal surface excitons. For clarity, we focus on Device 2's EL behavior in the main text, with Device 1 (thicker CrSBr) data in the Supplementary Information.

Obvious EL from Device 2 emerged at applied voltages exceeding 2.5 V (0.5 $\upmu$A), characterized by a dominant surface excitonic P-peak accompanied by weaker $\mathrm{X}_d$ and $\mathrm{X}^*$ features (Figure \ref{F2}a). Increasing bias voltages promoted exciton generation and consequent radiative recombination in CrSBr through inelastic tunneling processes. The EL spectra displayed progressively broader linewidths at elevated biases, a phenomenon likely attributed to increased thermal effects within the device. Additionally, the asymmetric line shape of the excitonic emission suggests the involvement of phonon-assisted indirect exciton transitions, consistent with the nearly flat conduction band dispersion along the $\Gamma$–X direction observed in CrSBr\cite{2021phonon}. Notably, neither device exhibited detectable EL signals under reverse bias conditions.

\begin{figure*}[tbh]    
\includegraphics[width=1.7\columnwidth]{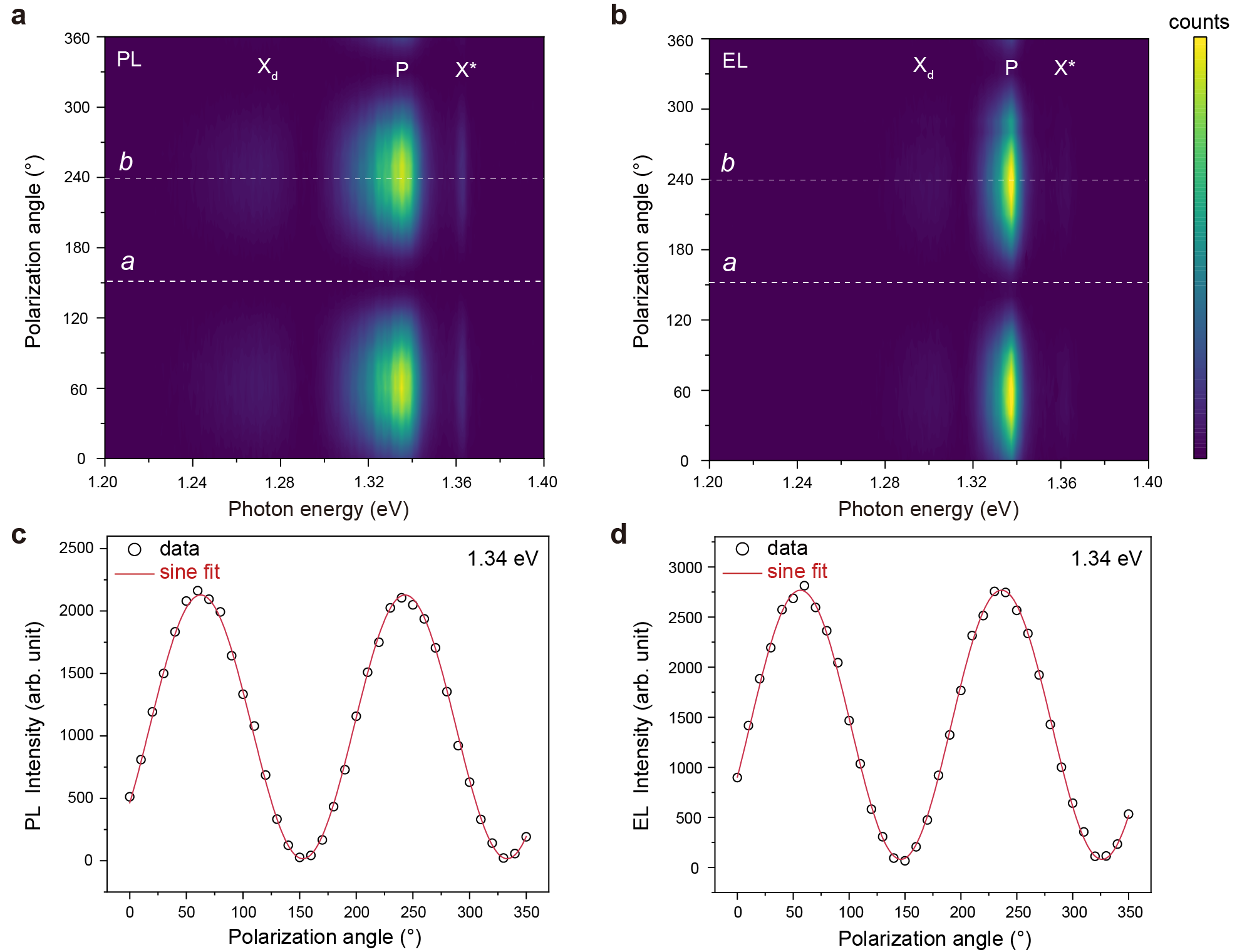}
\captionsetup{singlelinecheck=off, justification = RaggedRight}
\caption{\label{F3}\textbf{Polarized light emission of PL and EL for Device 2.}
     \textbf{(a,b)} Polarized PL and EL (under an applied bias of 3.1 V) spectra at different detection angles at 2 K. The intensity maximizes when the detection angle is parallel to the $b$-axis.  
     \textbf{(c,d)} PL/EL intensity versus the detection angles at 1.34 eV. The red line is a trigonometric fitting of data. The DOLP of the P peak is 98.7\% (94.3\%) for PL (EL). The angle of the polarizer is an arbitrary angle set in these measurements.
}
\end{figure*}

Remarkably, our device exhibited photon emission at ultralow currents (\textless 1 $\upmu$A), while conventional 2D \textit{p}-\textit{n} diodes\cite{2014pn} or metal-insulator-semiconductor (MIS) architectures\cite{2017MIS} typically require much higher currents for observable EL. This enhanced efficiency stems from the inelastic tunneling mechanism, as evidenced by the linear correlation between EL intensity and tunneling current (Figure \ref{F2}b), indicating nearly constant quantum efficiency across excitations. This behavior originates from the energy transfer mechanism that bypasses the need for direct carrier injection into the emission layer. To elucidate the temperature dependence of EL characteristic, we performed temperature-dependent measurements on Device 2 (Figure \ref{F2}c). The normalized EL spectra revealed a distinct redshift of the P peak between 2 K and 90 K, in agreement with the Varshni relation\cite{1967Varshni}, which accounts for band structure renormalization through lattice constant variation and electron-phonon interaction. Concurrently, we observed significant peak broadening at elevated temperatures. Complete emission quenching occurred above 110 K, where non-radiative recombination pathways became dominant. The maintained EL response across temperatures below 110 K and varying bias voltages (Figure S3) further demonstrates the robust thermal stability of Device 2.

The polarization characteristics of both PL and EL emissions were systematically investigated (Figure \ref{F3}). The PL intensity showed strong in-plane anisotropy, with maximum and minimum emission intensity corresponding to the crystallographic $b$- and $a$-axes, respectively (Figure \ref{F3}a), in agreement with previous studies\cite{wilson2021interlayer}. Remarkably, the EL maintained identical polarization alignment (Figure \ref{F3}b), demonstrating consistent anisotropic emission properties. Quantitative analysis of the 1.34 eV excitonic peak (P) revealed exceptional linear polarization, with a DOLP (defined as $(I_{\rm{max}}-I_{\rm{min}})/(I_{\rm{max}}+I_{\rm{min}})$) values of 98.7\% (PL, Figure \ref{F3}c) and 94.3\% (EL, Figure \ref{F3}d), significantly exceeding those reported polarized light-emitting diodes (LEDs) based on anitropic 2D semiconductors, such as $\mathrm{ReS_2}$\cite{2020Res2}. Device 1, incorporating thicker CrSBr, exhibited even higher EL DOLP (99.0\% at 1.34 eV, Figure S4), further confirming the robust polarization characteristic of this material system. 

This robust polarization stems from CrSBr's quasi-1D excitonic nature, arising from the intricate interplay of effective mass anisotropy and dielectric anisotropy\cite{2023bulk-dft}. Density functional theory (DFT) calculations reveal a nearly flat conduction band along $\Gamma$-X ($a$-axis) versus dispersive $\Gamma$-Y ($b$-axis) directions\cite{2023bulk-dft,2023charge-DFT}, enabling strongly polarized optical transitions. The preserved EL polarization confirms that inelastic tunneling-mediated energy transfer maintains intrinsic anisotropic exciton properties without depolarization. Such near-unity EL linear polarization in the near-infrared range highlights CrSBr's potential for on-chip polarized optoelectronics.

\begin{figure*}[tbh]    
\includegraphics[width=2\columnwidth]{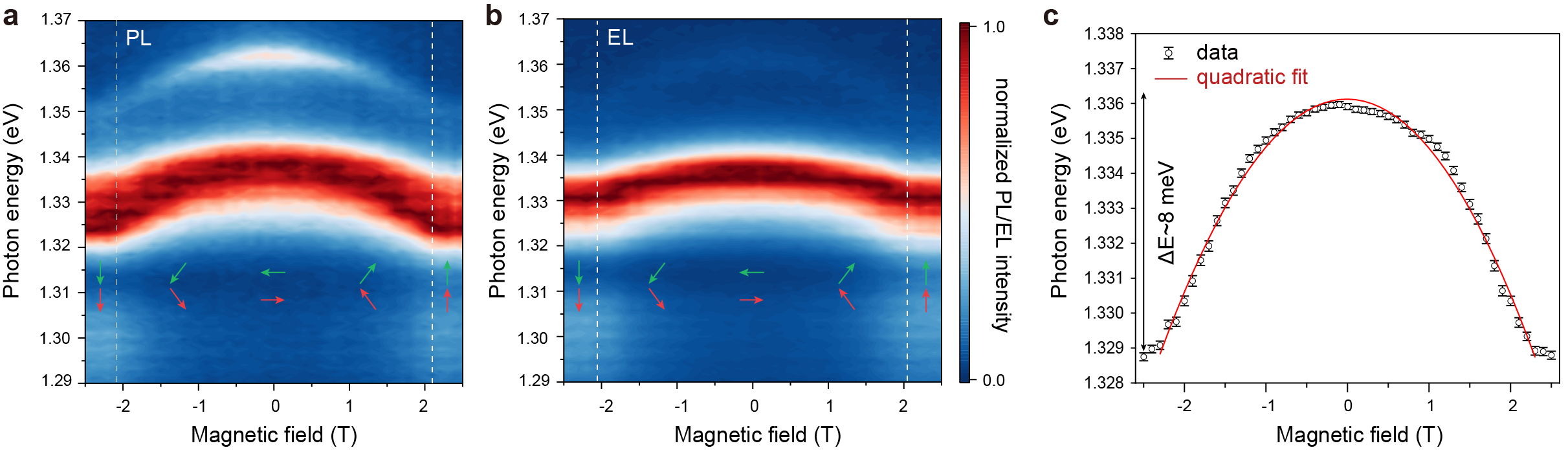}
\captionsetup{singlelinecheck=off, justification = RaggedRight}
\caption{\label{F4}\textbf{Magnetic-field-induced wavelength-tunable PL and EL emission.}
     \textbf {(a,b)} Magnetic field dependence of normalized PL and EL ($V_{\rm{bias}}$=3.1 V) spectra at 2 K with the field applied along the hard $c$-axis. Colored arrows illustrate the magnetization orientation in adjacent layers: AFM at $B$=0 T, canted spins at the intermediate field, and FM alignment about the saturation field ($B_{\rm{sat}}$). The P peak exhibits a continuous redshift up to $B_{\rm{sat}}$. 
     \textbf{(c)} EL peak position of P peak as a function of the magnetic field. The red curve shows a quadratic fit ($\Delta E \propto B^2$) to the experimental data below $B_{\rm{sat}}$, with peak shifting by 8 meV.
}
\end{figure*}

The strong magneto-excitonic coupling in CrSBr enables unprecedented magnetic control of EL—a critical feature for developing tunable on-chip light sources. This unique capability stems from CrSBr's A-type antiferromagnetic (AFM) order, where spin canting under external fields directly modifies excitonic interactions. As shown in Figure \ref{F4}a, applying a magnetic field along the hard $c$-axis induces a continuous redshift of the P peak in PL spectra up to the saturation field $B_{\rm{sat}}=\pm$2.3 T, beyond which the energy stabilizes\cite{2023nature_magneto-optics,2024spin-exciton,2023NNexcitons-magnons,2022nature-exciton-magnon}. Remarkably, EL spectra under 3.1 V bias exhibit identical tuning behavior (Figure \ref{F4}b), with an 8 meV redshift as the interlayer configuration transitions from  AFM to ferromagnetic (FM) via spin canting. The parallel response of the $\mathrm{X}^*$ peak confirms its shared origin with P, while the invariant defect-related $\mathrm{X}_d$ peak demonstrates its spin-decoupled nature. Similar magnetic-field-dependent behavior is observed in Device 1, as shown in Figure S6. Notably, the high DOLP (94.8\% at 1.33 eV under 2.5 T; Figure S5) persists during tuning, preserving polarization control. 

The energy shift arises from field-dependent interlayer exciton hybridization. In the AFM state, spin antiparrallelism decouples adjacent layers, whereas FM alignment enables resonant interlayer coupling that redistributes exciton wavefunctions and binding energies. Quantitative analysis reveals a quadratic energy shift of the P peak $\Delta E \propto B^2$ below $B_{\rm{sat}}$ (Figure \ref{F4}c), governed by spin-alignment-modulated interlayer hopping $t_h \propto ⟨S_1|S_2⟩ \propto cos(\theta/2)$, where $\theta$ represents the angle of magnetization direction between adjacent layers\cite{wilson2021interlayer}. The resultant bandgap renormalization highlights CrSBr's singular capacity to bridge magnetic and photonic control in LEDs.

In conclusion, our work demonstrates a transformative approach to polarized and spectrally tunable EL through the 2D magnet CrSBr, which simultaneously achieves three breakthroughs: intrinsic near-unity linear polarization (DOLP $\approx$ 94.3$\%$) via quasi-1D excitons, continuous 8 meV magnetic tuning through spin-canting control, and ultralow-threshold operation (\textless 1 $\upmu$A) enabled by efficient exciton-assisted tunneling. By leveraging CrSBr’s unique magneto-excitonic coupling, we overcome the traditional trade-offs between polarization purity, spectral tunability, and efficiency that plague conventional semiconductor devices. This platform establishes magnetic vdW materials as a versatile interface for coordinating spin, charge, and photon interactions—enabling compact, multifunctional light sources for applications ranging from real-time polarization spectroscopy to reconfigurable optical interconnects.

\bigskip

\bigskip
\noindent 
\textbf{Methods}\\
\noindent \textbf{Crystal synthesis and device fabrication.} 
CrSBr single crystals were grown via the chemical vapor transport method. A mixture of disulfur dibromide and chromium metal in a 7:13 molar ratio was sealed in a vacuum quartz tube. The sealed tube was then placed in a two-zone tube furnace, where CrSBr crystals were grown under a temperature gradient from 950 °C to 880 °C over seven days, with a heating/cooling rate of 1 °C/min. Graphite was employed to minimize interfacial roughness and facilitate efficient carrier injection. All flakes were mechanically exfoliated onto Si/$\mathrm{SiO_2}$ substrates and subsequently assembled layer by layer using a dry transfer technique onto prepatterned electrodes. The sample thickness was characterized by atomic force microscopy.

\noindent \textbf{Optical Measurements.} 
Optical measurements were performed using a closed-cycle helium cryostat (attoDRY2100) with a base temperature of 1.6 K and an out-of-plane magnetic field capability of up to 9 T. PL measurements were carried out using a HeNe laser (633 nm), which was focused onto the sample through a high-numerical-aperture (0.82) objective, producing a beam spot approximately 1 $\upmu$m in diameter. The emitted light was collected by a spectrometer (SpectraPro HRS-500S) and detected with a liquid-nitrogen-cooled charge-coupled device (PyLoN:400). EL signals of Device 1 (Device 2) was collected using a grating with a groove density of 150 g/mm, integrating 60 s (5 s). Spatial mapping of the EL signal was accomplished by controlling the piezoelectric sample stage.

\bigskip
\noindent\textbf{Data availability}\\
\noindent
All relevant data are available in the main text, Supporting Information, or upon request to the authors.

\bigskip
\noindent\textbf{Acknowledgment}\\
\noindent
This work was supported by the National Natural Science Foundation of China (Grants No. 12425402 and No. 12250007), the National Key R\&D Program of China (Grants No. 2022YFA1203902 and No. 2023YFF1500600), and the China Postdoctoral Science Foundation (2023TQ0003 and 2023M740122). \\

\bigskip
\noindent\textbf{Author contributions}\\
\noindent
Y.Y., S.Y., and Y.W. conceived the project, designed the experiments, analyzed the results, and wrote the manuscript. Y.W., L.H., and Y.R. fabricated the device and performed the measurements. Z.C. grew the CrSBr single crystal. K.W. and T.T. grew the \textit{h}-BN single crystals. \\

\bigskip
\noindent\textbf{Competing interests}\\
\noindent
During the preparation of this manuscript, the authors became aware of parallel findings posted in preprints by other research groups \cite{2025arxiv}.\\

\bigskip
\noindent\textbf{Additional information}\\
\noindent
\textbf{Supporting information} The online version contains Supporting material available at URL.\\

\normalem
\bibliographystyle{naturemag}
\bibliography{main}

\end{document}